\documentclass[fleqn,twoside]{article} 
\usepackage{epsf,multicol,ifthen}
\usepackage{ujp}
\usepackage[cp1251]{inputenc}
\usepackage[english,russian]{babel}
\usepackage{amstext}
\nazva{The Bose-Einstein correlations from a Quantum Field Theory
perspective}%

\udk{preprint }

\nazvacol{The Bose-Einstein correlations from a
QFT ... }%

\avtor{G.A.Kozlov$\footnotesize^1$, O.V.Utyuzh$\footnotesize^2$ and G.Wilk$\footnotesize^2$}%
\avtorcol{G.A.Kozlov, O.V.Utyuzh and G.Wilk }%

\inst{Bogoliubov Laboratory of Theoretical Physics, Joint Institute for Nuclear Research}%
\adr{(141980 Dubna,Moscow Region, Russia), }
\insti{The Andrzej So\l tan Institute for Nuclear Studies}%
\adri{(Ho\.za 69; 00-689 Warsaw, Poland) }%

\begin{document}           
\setcounter{page}{1}%
\maketitl                 
\begin{multicols}{2}
\anot{%
Using  a specific version of thermal Quantum Field Theory (QFT),
supplemented by operator-field evolution of the Langevin type,
we discuss two issues concerning the Bose Einstein correlations
(BEC): the origin of different possible coherent behaviour of the
emitting source and the origin of the observed shape of the BEC
function $C_2(Q)$. We demonstrate that previous conjectures in this
matter obtained by other approaches are confirmed and have received
complementary explanation. 
}%

\section{Introduction}

The Bose-Einstein correlations (BEC) are since long time recognized
as very important tool providing informationТs about hadronization
processes not available otherwise, especially in what concerns
space-time extensions and coherent or chaotic character of the
hadronizing sources. Because the importance of BEC and their present
experimental and theoretical status are widely known and well
documented (see, for example, \cite{BEC} and references therein), we
shall not repeat it here. Instead we shall proceed to main point of
our interest here, already mentioned above, namely to discussion of:
$(i)$ how the possible coherence of the hadronizing system influences
the two body BEC function $C_2(Q)$ \cite{Weiner,ALS},
\begin{equation}
C_2(Q) = \frac{N_2(k,k')}{N_1(k)N_1(k')} \label{eq:defC2}
\end{equation}
and $(ii)$ what is the true origin of the experimentally observed
$Q$-dependence of the $C_2(Q)$ correlation function in the approach
used here (out of which the space time information is being deduced,
$Q=|k_{\mu}-k'_{\mu}| = \sqrt{(k_{\mu}- k'_{\mu})^2}$ with
$k_{\mu}$ and $k'_{\mu}$ being the four-momenta of detected
particles; in what follows we shall for simplicity assume that all
produced particles are bosons). In literature one finds that in some
approaches using quantum statistical methods \cite{Weiner}
\begin{equation}
C_2(Q) = 1 + 2p(1-p)\cdot \sqrt{\Omega(Q\cdot r)}\, +\, p^2\cdot \Omega(Q\cdot r) ,
\label{eq:C2final} 
\end{equation}
whereas in other approaches \cite{ALS}
\begin{equation}
C_2(Q) = 1 + \lambda\cdot \Omega(Q\cdot r), \label{eq:usual}
\end{equation}
(actually, this is the most frequently used form). In both cases
parameters $p$ and $\lambda$ are called {\it coherence} parameters
defining the degree of coherence of hadronizing source (for purely
coherent source $p=\lambda=0$ and there is no BEC, for $p=\lambda
=1$, i.e., in purely chaotic case, both equations coincide). Although
in \cite{Weiner,ALS} they are operationally expressed in the same
way, i.e., $p=\lambda = \langle N_{chaotic}\rangle/\langle
N_{total}\rangle$, we shall formally differentiate between them when 
using the corresponding expressions for $C_2(Q)$ because, as will be
shown later, the concepts of coherence they correspond to are
different in each case. The choices of $\Omega (Q)$ discussed in
literature vary between \cite{Sources,Podg,Pratt} (here $q=r\cdot Q$
with $r = |r_{\mu}| = \sqrt{r_{\mu}r_{\mu}}$ being a $4$-vector such
that $\sqrt{(r_{\mu})^2}$ has dimension of length and product $Q\cdot
r = Q_{\mu} r_{\mu}= q$ is dimensionless): 
\begin{itemize}
\item Gaussian: $\Omega(q) = \exp \left( - Q^2 r^2\right)$;
\item exponential: $\Omega(q) = \exp ( - Q r)$;
\item Lorentzian: $\Omega(q) = 1/\left(1 + Q r\right)^2$;
\item given by Bessel function \cite{Podg}:\\
  $\Omega(q) = \left[J_1(Q r)/(Q r)\right]^2$ .
\end{itemize}
The most frequently used forma are Gaussian and exponential ones
\cite{BEC}. The above questions can be therefore rephrased in the
following way: $(i)$ why are forms of $C_2(Q)$ in eqs.
(\ref{eq:C2final}) and (\ref{eq:usual}) different and are parameters
$p$ and $\lambda$ referring to the same quantity and $(ii)$ what
stays behind specific choice of the form of $\Omega (Q)$ function as
listed above.

To make our point more clear we shall work here directly in phase
space (as, for example, in \cite{Takagi}), no space-time
considerations will be used (contrary to the majority of works on BEC
\cite{BEC,Weiner,ALS}). As our working tool we shall choose   
some specific (thermal) version of Quantum Field Theory (QFT)
supplemented by the operator-field evolution of Langevin type
proposed recently \cite{Namiki,Kozlov,Langevin}. We shall demonstrate that: 
$(i)$ the origin of differences in $C_2(Q)$ in eqs.
(\ref{eq:C2final}) and (\ref{eq:usual}) lies in different ways of
introducing the concept of coherence in both approaches, i.e., $p$
and $\lambda$ referred to different concepts of coherence in each case;
$(ii)$ in order to obtain a given (experimentally observed) shape of
the BEC correlation function $C_2(Q)$ (i.e., the $\Omega (Q)$) one
has to account somehow for  the {\it finiteness} of the space-time
region of the particle production (i.e., of the hadronizing {\it
source}). In QFT approach used here it is particularly clearly seen
and is connected with the necessity of smearing out of some
generalized functions (delta functions:
$\delta(Q_{\mu}=k_{\mu}-k'_{\mu})$) appearing in the definition of
thermal averages of some operators occurring here. The freedom in
using different types of smearing functions to perform such a
procedure allows us to account for all possible different shapes of
hadronizing sources apparently observed by experiment. (Actually,
careful inspection of all previous approaches to BEC using QFT, cf.,
for example, \cite{RW}, shows that this was always the procedure
used, though never expressed so explicitly as is done here.).

\section{Description of hadronizing source}

Let us recapitulate now the main points of our approach (for details
see \cite{Kozlov}). The collision process produces usually a large
number of particles out of which we select one (we assume for
simplicity that we are dealing only with identical bosons) and
describe it by operator $b(\vec{k},t)$ (the notation is the usual
one: $b(\vec{k},t)$ is an annihilation operator, $\vec{k}$ is
$3$-momentum and $t$ is a real time). The rest of the particles are
then assumed to form a kind of heat bath, which remains in
equilibrium characterized by a temperature $T=1/\beta$ (which will be
one of our parameters). All averages $\langle (\dots) \rangle$ are
therefore thermal averages of the type:
\begin{equation}
\langle (\dots )\rangle = Tr\left[(\dots) e^{-\beta
H}\right]/Tr\left( e^{-\beta H}\right). \label{eq:Thav}
\end{equation}  
However, we shall also allow for some external (to the above heat
bath) influence acting on our system. Therefore we shall represent
the operator $b(\vec{k},t)$ as consisting of a part corresponding to
the action of the heat bath, $a(\vec{k},t)$, and also of a part
describing action of these external factors, $R(\vec{k},t)$:
\begin{equation}
b(\vec{k},t) = a(\vec{k},t) + R(\vec{k},t).\label{eq:apR}
\end{equation}
The time evolution of such a system is then assumed to be given by a
Langevin equation \cite{Langevin}
\begin{equation}
i\partial_t b(\vec{k},t) = F(\vec{k},t) - A(\vec{k},t) + P
\label{eq:Lang}
\end{equation}
(and a similar conjugate equation for $b^{+}(\vec{k},t)$). These
equations are supposed to model all aspects of the hadronization
process (notice their similarity to equations describing motion of
Brown particle in some external field \cite{Namiki}). The meaning of
different terms appearing here is following:\\ 
{\bf (i)} The combination $F(\vec{k},t)-A(\vec{k},t)$ represents the 
so called {\it Langevin force} and is therefore responsible for the
internal dynamics of hadronization in the following manner: 
\begin{itemize}
\item
$A$ is related to stochastic dissipative forces and is given by
\cite{Langevin,Kozlov} 
\begin{equation}
A(\vec{k},t) = \int^{+\infty}_{-\infty}\! d\tau K(\vec{k},t-\tau)
b(\vec{k},\tau), \label{eq:AL}
\end{equation}
with the operator $K(\vec{k},t)$ being a random evolution field
operator describing the random noise and satisfying the usual
correlation-fluctuation relation for the Gaussian noise:
\begin{equation}
\langle K^{+}(\vec{k},t)K(\vec{k}',t)\rangle = 2\sqrt{\pi \rho}\kappa
\delta(\vec{k}-\vec{k}') \label{eq:Gn}
\end{equation}
($\kappa$ and $\rho$ are parameters describing effect caused by this
noise on the particle evolution in thermal environment
\cite{Langevin}). 
\item
The operator $F(\vec{k},t)$ describes the influence of heat bath,
\begin{equation}
F(\vec{k},t) =
\int^{+\infty}_{-\infty}\!\frac{d\omega}{2\pi}\psi(k_{\mu})\hat{c}(k_{\mu})
e^{-i\omega t} . \label{eq:FL}
\end{equation}
\end{itemize}
{\bf (ii)} Our heat bath is represented by an ensemble of damped oscillators,
each described by operator $\hat{c}(k_{\mu})$ such
that $\left[\hat{c}(k_{\mu}),\hat{c}^{+}(k'_{\mu})\right] =
\delta^4(k_{\mu}-k'_{\mu})$, and characterized by some function
$\psi(k_{\mu})$, which is subjected to is a kind of normalization
involving also dissipative forces represented by Fourier transformed
operator $\tilde{K}(k_{\mu})$, namely:
\begin{equation}
\int^{+\infty}_{-\infty}\frac{d\omega}{2\pi} 
                \left[\frac{\psi(k_{\mu})}
                {\tilde{K}(k_{\mu})-\omega}\right]^2= 1 .\label{eq:Psi}
\end{equation}
{\bf (iii)} Finally, the constant term $P$ (representing {\it external source}
term in Langevin equation) denotes the possible influence of
some external force (assumed here to be constant in time). This force
would result, for example, in a strong ordering of phases, leading 
therefore to the coherence effect in the sense discussed in
\cite{Weiner}. 

Out of many details (for which we refer to \cite{Kozlov}) what is
important in our case is the fact that the $2$-particle correlation
function for like-charge particles, as defined in (\ref{eq:defC2}),
is given in such form ($(k_{\mu} = (\omega=k^{0},k_j)$):
\begin{eqnarray}
C_2(Q) &=&  \xi(N) \cdot \frac{\tilde{f}(k_{\mu},k'_{\mu})}
{\tilde{f}(k_{\mu})\cdot\tilde{f}(k'_{\mu})}\nonumber\\
&=& \xi(N) \cdot \left[1\, +\, D(k_{\mu},k'_{\mu})\right] ,
\label{eq:C2}
\end{eqnarray}
where 
\begin{eqnarray}
\tilde{f}(k_{\mu},k'_{\mu}) &=& \langle
\tilde{b}^{+}(k_{\mu})\tilde{b}^{+}(k'_{\mu})
\tilde{b}(k_{\mu})\tilde{b}(k'_{\mu})\rangle ,\nonumber \\
\tilde{f}(k_{\mu}) &=& \langle
\tilde{b}^{+}(k_{\mu})\tilde{b}(k_{\mu})\rangle \label{eq:fff}
\end{eqnarray}
are the corresponding thermal statistical averages (in which
temperature $T$ enters as a parameter) with $\tilde{b}(k_{\mu}) =
\tilde{a}(k_{\mu}) + \tilde{R}(k_{\mu})$
being the corresponding Fourier transformed stationary solution
of eq. (\ref{eq:Lang}). As shown in \cite{Kozlov} (notice that
operators $\tilde{R}(k_{\mu})$ by definition commute with themselves
and with any other operator considered here):
\begin{eqnarray}
\tilde{f}(k_{\mu},k'_{\mu}) &=& \tilde{f}(k_{\mu})\cdot\tilde{f}(k'_{\mu}) +\nonumber\\
&+& \langle\tilde{a}^{+}(k_{\mu})\tilde{a}(k'_{\mu})\rangle
\langle\tilde{a}^{+}(k'_{\mu})\tilde{a}(k_{\mu})\rangle  + \nonumber\\
&+&\langle\tilde{a}^{+}(k_{\mu})\tilde{a}(k'_{\mu})\rangle \tilde{R}^{+}(k'_{\mu})\tilde{R}(k_{\mu}) +\nonumber\\
&+&\langle\tilde{a}^{+}(k'_{\mu})\tilde{a}(k_{\mu})\rangle \tilde{R}^{+}(k_{\mu})\tilde{R}(k'_{\mu}) ,\label{eq:ff}\\
\tilde{f}(k_{\mu}) &=& \langle \tilde{a}^+(k_{\mu})\tilde{a}(k_{\mu})\rangle\, +\,
|\tilde{R}(k_{\mu})|^2 . \label{eq:f}
\end{eqnarray}
This defines $D(k_{\mu},k'_{\mu})$ in (\ref{eq:C2}) in terms of the
operators $\tilde{a}(k_{\mu})$ and $\tilde{R}(k_{\mu})$, which in our
case are equal to: 
\begin{equation}
\tilde{a}(k_{\mu}) =
\frac{\tilde{F}(k_{\mu})}{\tilde{K}(k_{\mu}) - \omega}\quad {\rm and}\quad
\tilde{R}(k_{\mu}) =
\frac{P}{\tilde{K}(k_{\mu}) - \omega} . \label{eq:aR}
\end{equation}
The multiplicity $N$ depending factor $\xi$ is in our case equal to 
$\xi (N) = \langle N\rangle^2/\langle N(N-1)\rangle$. This means therefore
that the correlation function $C_2(Q)$, as defined by eq.
(\ref{eq:C2}), is essentially given in terms of $P$ and the two
following thermal averages for the $F(\vec{k},t)$ operators:
\begin{eqnarray}
\langle F^{+}(\vec{k},t)F(\vec{k}',t')\rangle &=&
\delta^3(\vec{k}-\vec{k}')\cdot\nonumber\\
&\cdot&\int \frac{d\omega}{2\pi}\,
               \left|\psi\right|^2\, n(\omega)e^{+i\omega(t-t')},
               \label{eq:theavc}\\
\langle F(\vec{k},t)F^{+}(\vec{k}',t')\rangle &=&
\delta^3(\vec{k}-\vec{k}')\cdot \nonumber\\
&\cdot&\int \frac{d\omega}{2\pi}\,
               \left|\psi\right|^2\, [1 + n(\omega)]
e^{-i\omega(t-t')}\nonumber
\end{eqnarray}
where $n(\omega) = \left\{\exp \left[(\omega - \mu)\beta\right] -
1\right\}^{-1}$ is the number of (by assumption - only bosonic in our
case) damped oscillators of energy $\omega$ in our reservoir
characterized by parameters $\mu$ (chemical potential) and inverse
temperature $\beta=1/T$ (both being free parameters). The origin of
these parameters, the temperature $\beta =1/T$ and chemical potential
$\mu$, is the Kubo-Martin-Schwinger condition that 
\begin{equation}
\langle a(\vec{k}',t')a^{+}(\vec{k},t)\rangle = \langle
a^{+}(\vec{k},t)a({k}',t-i\beta)\rangle \cdot \exp(-\beta \mu), \label{eq:KMS}
\end{equation}
(see \cite{Kozlov,Langevin}). This form of averages presented in
(\ref{eq:theavc}) reflects the corresponding averages for the
$\hat{c}(k_{\mu})$ operators, namely that 
\begin{eqnarray}
\langle \hat{c}^{+}(k_{\mu})\hat{c}(k'_{\mu})\rangle 
                 &=& \delta^4(k_{\mu} - k'_{\mu})\cdot n(\omega)\nonumber\\
\langle \hat{c}(k_{\mu})\hat{c}^{+}(k'_{\mu})\rangle 
                 &=&  \delta^4(k_{\mu} - k'_{\mu})\cdot [1 +
n(\omega)] .\label{eq:coper}
\end{eqnarray}
Notice that with only delta functions present
in (\ref{eq:theavc}) one would have a situation in which our hadronizing
system would be described by some kind of {\it white noise} only. The
integrals multiplying these delta functions and depending on $(a)$ momentum
characteristic of our heat bath $\psi(k_{\mu})$ (representing in our case,
by definition, the hadronizing system) and $(b)$ assumed bosonic statistics
of produced secondaries resulting in factors $n(\omega)$ and $1+n(\omega)$,
respectively, bring the description of our system closer to
reality.

\section{Results}

It is straightforward to realize that the existence of BEC, i.e.,
the fact that $C_2(Q) >1$, is strictly connected with nonzero values
of the thermal averages (\ref{eq:theavc}). However, in the form
presented there, they differ from zero {\it only at one point},
namely for $Q=0$ (i.e., for $k_{\mu} = k'_{\mu}$). Actually, this is
the price one pays for the QFT assumptions tacitly made here, namely
for the {\it infinite} spatial extension and for the {\it uniformity}
of our reservoir. But we know from the experiment \cite{BEC} that
$C_2(Q)$ reaches its maximum at $Q=0$ and falls down towards its
asymptotic value of $C_2 = 1$ at large of $Q$ (actually already at $Q
\sim 1$ GeV/c). To reproduce the same behaviour by means of our
approach here, one has to replace delta functions in eq.
(\ref{eq:theavc}) by functions with supports larger than limited to a
one point only. This means that such functions should not be infinite
at $Q_{\mu} = k_{\mu}-k'_{\mu} =0$ but remain more or less sharply
peaked at this point, otherwise remaining finite and falling to zero
at small, but finite, values of $|Q_{\mu}|$ (actually the same as
those at which $C_2(Q)$ reaches unity): 
\begin{equation}
\delta(k_{\mu} - k'_{\mu})\, \Longrightarrow\, \Omega_0\cdot
\sqrt{\Omega(q=Q\cdot r)}.
\label{eq:Om}
\end{equation}
Here $\Omega_0$ has the same dimension as the $\delta$ function
(actually, it is nothing else but $4$-dimensional volume restricting
the space-time region of particle production) and $\Omega(q)$ is a
dimensionless smearing function which contains the $q$-dependence we
shall be interested in here.  In this way we are tacitly introducing
a new parameter (mentioned already in the Introduction), $r_{\mu}$, a
$4$-vector such that $\sqrt{(r_{\mu})^2}$ has dimension of length and
which makes the product $Q\cdot r = Q_{\mu} r_{\mu}= q$
dimensionless. This defines the region of {\it nonvanishing} density
of oscillators $\hat{c}$, which we shall {\it identify} with the
space-time extensions of the hadronizing source. The expression
(\ref{eq:Om}) has to be understood in a symbolic sense, i.e., that
$\Omega(Q\cdot r)$ is a function which in the limit of $r\rightarrow
\infty$ becomes {\it strictly} a $\delta$ function. Making such
replacement in eq. (\ref{eq:theavc}) one must also decide how to
accordingly adjust $n(\omega)$ occurring there because now, in
general, $\omega \neq \omega'$. In what follows we shall simply
replace $n(\omega) \rightarrow n(\bar{\omega})$ with $\bar{\omega} =
(\omega + \omega')/2$ (which, for classical particles would mean that
$n(\omega) \rightarrow \sqrt{n(\omega)n(\omega')}$). 

In such way $r$ becomes new (and from the phenomenological point of
view also the most important) parameter entering here together with
the whole function $\Omega(Q\cdot r)$, to be deduced from comparison
with experimental data (one should notice at this point that the
opposite line of reasoning has been used in \cite{Z} where at first a
kind of our $\Omega(q)$ function was constructed for a finite source
function and it was then demonstrated that in the limit of infinite,
homogenous source one ends with a delta function). With such a
replacement one now has 
\begin{equation}
D(k_{\mu},k'_{\mu}) =
\frac{\sqrt{\tilde{\Omega}(q)}}{(1+\alpha)(1+\alpha')}
\cdot \left[ \sqrt{\tilde{\Omega}(q)} + 2\sqrt{\alpha \alpha'}
\right]  \label{eq:res}
\end{equation}
where
\begin{equation}
\tilde{\Omega}(q) = \gamma \cdot \Omega(q),~~
\gamma = \frac{n^2(\bar{\omega})}{n(\omega)n(\omega')},~~
\alpha \propto \frac{P^2}{|\psi(k_{\mu})|^2 n(\omega)},
\label{eq:lambda}
\end{equation}
with $n(\omega)$ the same as defined above. 

Another very important parameter entering (\ref{eq:res}) is $\alpha$.
which first of all reflects action of external force $P$ present in
the evolution equation (\ref{eq:Lang}). This action is combined here
(in a multiplicative way) with information on both the the momentum
dependence of the reservoir (via $|\psi(k_{\mu})|^2$) and on the
single particle distributions of the produced particles (via
$n(\omega = \mu_T \cosh y)$ where $\mu_T$ and $y$ are, respectively,
the transverse mass and rapidity). Parameter $\alpha$ summarizes
therefore our knowledge of other than space-time characteristics of
the hadronizing source (given by $\Omega(q)$ introduced above).
Notice that $\alpha > 0$ only when $P \neq 0$. Actually, for $\alpha
= 0$ one has 
\begin{equation}
1 < C_2(Q) < 1 + \gamma \Omega(Q\cdot r) ,
\label{eq:limits}
\end{equation}
i.e., it is contained between limits corresponding to very large
(lower limit) and very small (upper limit) values of $P$.
Because of this $\alpha$ plays the role of the {\it coherence}
parameter \cite{BEC,Weiner}. For $\gamma \simeq 1$, neglecting
the possible energy-momentum dependence of $\alpha$ and assuming that
$\alpha' = \alpha$ one gets the expression
\begin{equation}
C_2(Q) = 1 + \frac{2\alpha}{(1 + \alpha)^2}\cdot
\sqrt{\Omega(q)}\, +\, \frac{1}{(1+\alpha)^2}\cdot
\Omega(q) , \label{eq:C2finala}
\end{equation}
which is formally {\it identical} with eq. (\ref{eq:C2final})
obtained in \cite{Weiner} by means of QS approach. It has precisely
the same form, consisting two $Q-$dependent terms containing the
information on the shape of the source, one being the square of the
other, each multiplied by some combination of the {\it chaoticity}
parameter $p = 1/(1+\alpha)$ (however, in \cite{Weiner} $p$ is
defined as the ratio of the mean multiplicity of particles produced
by the so called {\it chaotic} component of the source to the mean
total multiplicity, $p=\langle N_{ch}\rangle/ \langle N\rangle$). In
fact, because in general $\alpha \neq \alpha'$ (due to the fact that
$\omega \neq \omega'$ and therefore the number of states, identified
here with the number of particles with given energy, $n(\omega)$, are
also different) one should rather use the general form (\ref{eq:C2})
for $C_2$ with details given by (\ref{eq:res}) and (\ref{eq:lambda})
and with $\alpha$ depending on such characteristics of the production
process as temperature $T$ and chemical potential $\mu$ occurring in
definition of $n(\omega)$. 

Notice that eq. (\ref{eq:C2finala}) differs from the usual
empirical parameterization of $C_2(Q)$ \cite{BEC} as given by
eq.(\ref{eq:usual}) in which $0< \lambda <1$ is a free parameter
adjusting the observed value of $C_2(Q=0)$, which is customary called
"incoherence" and with $\Omega(Q\cdot r)$ represented usually as
Gaussian. Recently eq. (\ref{eq:usual}) has found strong theoretical
support expressed in great detail in \cite{ALS} (where $\lambda$ is
given by the same ratio od multiplicities as $p$ above). The natural
question arises: which of the two formulas presented here is correct?
The answer we propose is: both are right in their own way. This is
because each of them is based on different ways of defining coherence
of the source. In \cite{ALS} one uses the notion of coherently and
chaotically produced particles or, in other words, one divides
hadronizing source into coherent and chaotic subsources. In
\cite{Weiner} one introduces instead the notion of partially coherent
fields representing produced particles, i.e., one has only one
source, which produces partially coherent fields. Our approach is
similar as we describe our particle by operator $b(\vec{k},t)$, which
consists of two parts, cf. eq. (\ref{eq:apR}), one of which depends
on the external static force $P$. The action of this force is to {\it
order phases} of particles in our source (represented by the heat
bath). The strength of this ordering depends on the value of the
external force $P$. In any case, for $P \ne 0$, it demonstrates
itself as a {\it partial coherence}. 

Some comments are necessary at this point. Notice that it is of the
same type as that considered in \cite{Weiner}. When comparing with
\cite{ALS} one should notice that although our operators
$\tilde{a}(k_{\mu})$ and $\tilde{R}(k_{\mu})$ look similar to
operators defined in eqs. (4) and (5) of \cite{ALS} they differ in
the following. Our $R(k_{\mu})$ describes essentially the action of
{\it constant} force $P$ and as such it commutes with all other
operators (including themselves). So it only introduces a partial
ordering of phases of particles decreasing the $C_2$ correlation
function, i.e., acting as a coherent component, albeit we do not have
coherent particles as such. It is also seen when realizing that in
eq. (\ref{eq:f}) the two last terms contain only one pair of
operators $a$. This in the language of \cite{ALS} translates to only
one Wigner function, $f_{ch}$, to be present here. The operators $R$
cannot form the second Wigner function ($f_{coh}$ in \cite{ALS}).
This is the technical origin of the three terms present in
(\ref{eq:C2final}) (and in \cite{Weiner}) in comparison to two terms
in (\ref{eq:usual}) and obtained in \cite{ALS}.

Let us return to the problem of $Q$-dependence of BEC. One more
remark is in order here. The problem with the
$\delta(k_{\mu}-k'_{\mu})$ function encountered in two particle
distributions does not exist in the single particle distributions, 
which are in our case given by eq. (\ref{eq:f}) and which can be
written as $\tilde{f}(k_{\mu}) \propto \langle
\tilde{a}^+(k_{\mu})\tilde{a}(k_{\mu})\rangle\, +\,
|\tilde{R}(k_{\mu})|^2 \, \sim (1+\alpha)\langle
\tilde{a}^+(k_{\mu})\tilde{a}(k_{\mu})\rangle$ (it is normalized to 
the mean multiplicity: $\int d^4k\, \tilde{f}(k_{\mu}) = \langle
N\rangle$). To be more precise
\begin{equation}
\tilde{f}(k_{\mu})\, = \, (1+\alpha) \cdot \Xi(k_{\mu},k_{\mu}) ,
 \label{eq:Single}
\end{equation}
where $\Xi(k_{\mu},k_{\mu})$ is one-particle distribution function
for the "free" (undistorted) operator $\tilde{a}(k_{\mu})$ equal to
\begin{equation}
\Xi(k_{\mu},k_{\mu})\, =\, \Omega_0 \cdot \left| \frac{\psi(k_{\mu})}
                           {\tilde{K}(k_{\mu})-\omega}\right| ^2
n(\omega) . \label{eq:singlea}
\end{equation}
Notice that the actual shape of $\tilde{f}(k_{\mu})$ is dictated both
by $n(\omega) = n(\omega;T,\mu)$ (calculated for fixed temperature
$T$ and chemical potential $\mu$ at energy $\hat{\omega}$ as given by
the Fourier transform of random field operator $\tilde{K}$ and by
shape of the reservoir in the momentum space provided by
$\psi(k_{\mu})$) and by external force $P$ in parameter $\alpha$.
They are both unknown, but because these details do not enter the BEC
function $C_2(Q)$, we shall not pursue this problem further. What is
important for us at the moment is that both the coherent and the
incoherent part of the source have the same energy-momentum
dependence (whereas in other approaches mentioned here they were
usually assumed to be different). On the other hand it is clear from
(\ref{eq:Single}) that $\langle N\rangle = \langle N_{ch}\rangle +
\langle N_{coh}\rangle$ (where $\langle N_{ch}\rangle$ and $\langle
N_{coh}\rangle$ denote multiplicities of particles produced
chaotically and coherently, respectively) therefore justifying
definition of chaoticity $p$ mentioned above.

\noindent \epsfxsize=\columnwidth\epsffile{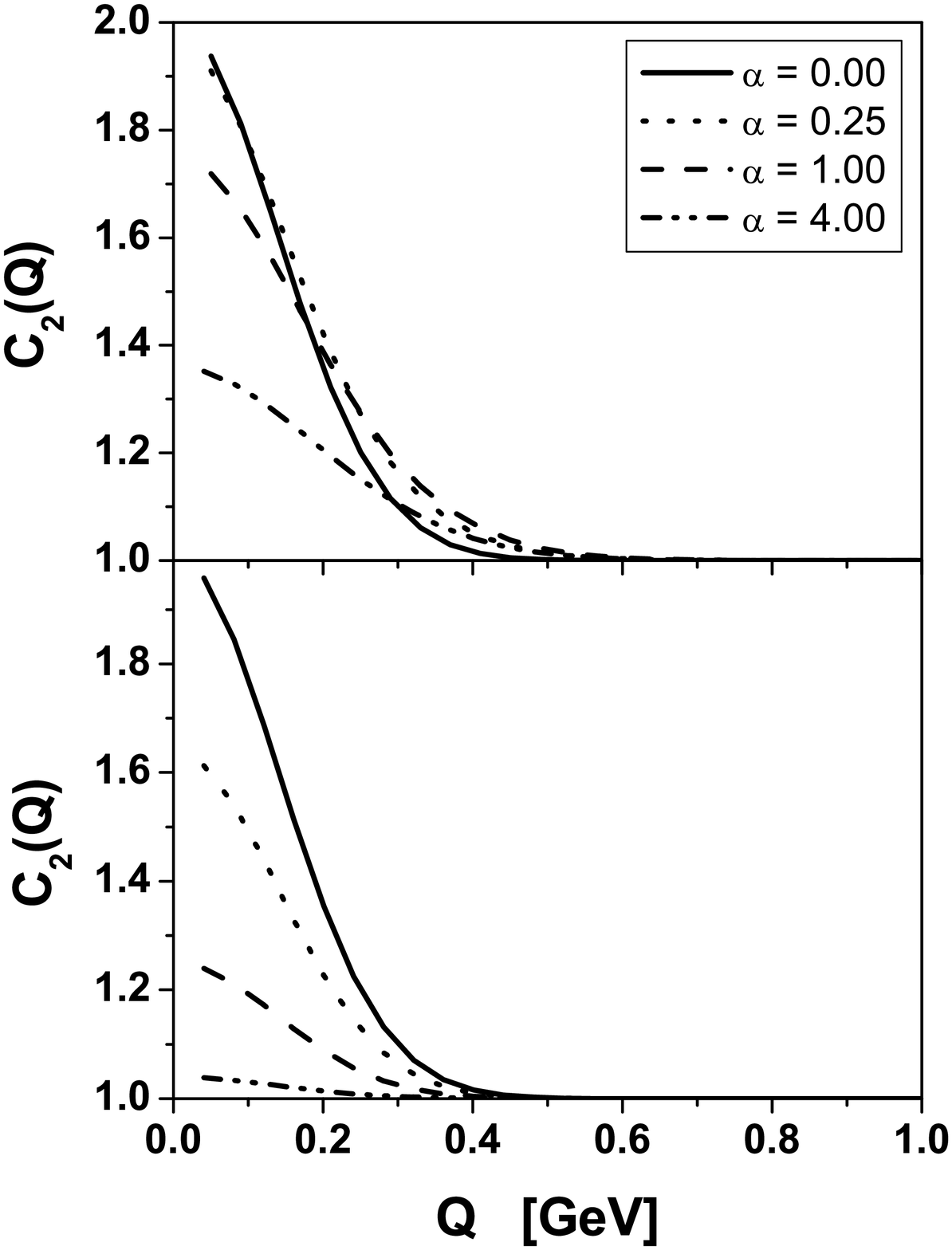}
\noindent{\footnotesize Figure 1. Shapes of $C_2(Q)$ as given by eq.
(\protect\ref{eq:C2final}) - upper panel and for the truncated
version of (\protect\ref{eq:C2final}) (without the middle term) 
corresponding to eq. (\ref{eq:usual}) - lower panel. Gaussian shape
of $\Omega(q)$ was used in both cases.}  

Fig. 1 shows in detail (using Gaussian shape of
$\Omega (q)$ function) the dependence of $C_2(Q)$ on different values of
$\alpha = 0,~0.25,~1,~4$ (again, used in the same approximate way as
before and corresponding to $p=1.,~0.8,~0.5,~0.2$) and compare it to
the case when the second term in eq. (\ref{eq:C2final}) is neglected,
as is the case in majority of phenomenological fits to data.

\section{Summary and conclusions}

To summarize: using a specific version of QFT supplemented by
Langevin evolution equation (\ref{eq:Lang}) to describe hadronization
process \cite{Kozlov,Langevin} we have derived the usual BEC
correlation function in the form explicitly showing: $(i)$ - the
origin of the so called coherence and its influences on the structure
of $C_2(Q)$; $(ii)$ - the origin of the $Q$-dependence of BEC
represented  by correlation function $C_2(Q)$. In our case the
dynamical source of coherence is identified with the existence of a
constant external term $P$ in the Langevin equation. Its influence
turns out to be identical with the one obtained before in the QS
approach \cite{Weiner} and is described by eq. (\ref{eq:res}). Its
action is to order phases of the produced secondaries: for
$P\rightarrow \infty$ all phases are aligned in the same way and
$C_2(Q) =1$. The coherence in \cite{Weiner} is thus property of
fields and in our case property of operators describing produced
particles. In both cases it occurs already on the level of a
hadronizing source. Dividing instead the hadronizing source 
itself into coherent and chaotic subsources results in eq.
(\ref{eq:usual}) obtained in \cite{ALS}. The controversy between
results given by \cite{Weiner} and \cite{ALS} is therefore resolved: 
both approaches are right, one should only remember that they use
different descriptions of the notion of coherence. It is therefore up
to the experiment to decide which proposition is followed by nature:
the simpler formula (\ref{eq:usual}) or rather the more involved
(\ref{eq:C2}) together with (\ref{eq:res}). From Fig. 1 one can see
that differences between both forms are clearly visible, especially
for larger values of coherence $\alpha$, i.e., for lower chaoticity
parameter $p$. 

From our presentation it is also clear that the form of $C_2$
reflects distributions of the space-time separation between the 
two observed particles rather than the distribution of their separate
production points (cf., for example, \cite{Zajc}, where it is
advocated that it is in fact a Fourier transform of two-particle
density profile of the hadronizing source,
$\rho(r_1,r_2)=\rho(r_1-r_2)$, without approximating it by the 
product of single-particle densities, as in \cite{BEC}).

Finally, we would like to stress that our discussion is so far
limited to only a single type of secondaries being produced. It is
also aimed at a description of hadronization understood as kinetic
freeze-out in some more detailed approaches. So far we were not
interested in the other (highly model dependent) details of the
particle production process. This is enough to obtain our general
goals, i.e., to explain the possible dynamical origin of coherence in
BEC and the origin of the specific shape of the correlation $C_2(Q)$
functions as seen from the QFT perspective. We close with suggestion
that both sources of coherence, that presented here and in
\cite{Weiner} and that investigated in \cite{ALS}, should be
considered together. The most general situation would be then
hadronizing source composed with a number of subsources, each with
different internal (discussed here) degree of coherence. For only one
subsource present we would have situation described here whereas for
a number of subsources, each being either totally coherent or totally
chaotic, description offered by \cite{ALS} would automatically
emerge. It is up to experimental data to decide and Fig. 2 tells us
that it is not impossible task (at least in principle).\\

We would like to acknowledge support obtained from the
Bogolyubov-Infeld program in JINR and partial support of the Polish
State Committee for Scientific Research (KBN): grants
621/E-78/SPUB/CERN/P-03/DZ4/99 and 2P03B05724.

\end{multicols}
\end{document}